# 6 BATCH INJECTION AND SLIPPED BEAM TUNE MEASUREMENTS IN FERMILAB'S MAIN INJECTOR

D. J. Scott, D. Capista, I. Kourbanis, K. Seiya, M.-J. Yan, Fermilab, Batavia, IL 60510, U.S.A.

*Abstract*

During NOVA operations it is planned to run the Fermilab Recycler in a 12 batch slip stacking mode. In preparation for this, measurements of the tune during a six batch injection and then as the beam is decelerated by changing the RF frequency have been carried out in the Main Injector. The coherent tune shifts due to the changing beam intensity were measured and compared well with the theoretically expected tune shift. The tune shifts due to changing RF frequency, required for slip stacking, also compare well with the linear theory, although some nonlinear affects are apparent at large frequency changes. These results give us confidence that the expected tunes shifts during 12 batch slip stacking Recycler operations can be accommodated.

## INTRODUCTION

The delivery of a high intensity proton beam for neutrino experiments is a core element of the Fermilab physics program for the next decade and beyond. Part of the accelerator upgrade portion of the NuMI Off-Axis νe Appearance Experiment, (NOVA) is to increase the beam power of the 120 GeV beam from the Main Injector (MI) synchrotron onto the neutrino target from the present maximum level of 400 kW to 700 kW. This increase of the proton throughput of the Main Injector (MI) will be achieved by converting the Recycler Ring (RR) from its function as an anti-proton storage ring during collider operations into a proton pre-injector for the MI [1].

In current operations the MI can accept up to 6 batches of 83 bunches from the Booster. The maximum intensity beam in the MI in current operations is achieved by slip-stacking two sets of 5 batches and then injecting a eleventh batch into the remaining gap. The beam is then accelerated to 120 GeV.

Using the RR as a pre-injector will enable the 12 batch slip-stacking to be performed in the RR, concurrently to the MI ramp, thereby reducing the MI cycle time from 2.2 s to 1.33 s, approximately doubling the proton throughput.

In preparation for this new operational mode of the RR studies in the MI were carried out to confirm the magnitude of the coherent tune shifts with increasing intensity and then as the beam is decelerated by reducing the RF frequency (required for slip stacking). These measurements helped confirm that there will be sufficient tune shift compensation in the RR.

## MEASUREMENTS

The beam intensity is proportional to the length of beam from the Linac that is injected into the Booster. Conveniently, this is measured in units of the length of a booster turn. Turn-by-turn (TBT) BPM data was taken for six injections of two to ten booster turns beam intensity. For the injection measurements the beam was kicked in the MI injection line to give a strong horizontal and vertical signal.

Turn by turn BPM data was also taken for the beam that was decelerated by the RF. For this data the beam was pinged with the injection kicker. Coupling was increased using a small time bump on a single skew quad. This would have a small effect on the intensity dependent tune shift that is assumed to be small. The chromaticity was set to -21 for all measurements to minimise losses.

## CALCULATING THE TUNE FROM THE BPM DATA

Data for each BPM were taken and analysed in the time and frequency domain to determine the tune, $Q_{x,y}$. An example calculation is described below in more detail.

*Frequency Domain*

Figure 1 shows a BPM signal and its Fourier Transform (FT). As there are approximately 200 BPMS in the MI and many measurements to consider an algorithm for finding the peak of the FFT was used to speed up the calculation. The FT was smoothed using a Gaussian filter and a moving average then the peak intensity found and this was assumed to be the tune. This method was applied to each BPM and there is a good agreement between BPMs for each measurement. Often when analysing the horizontal BPM data strong coupling to the vertical motion was seen, shown in Figure 2.

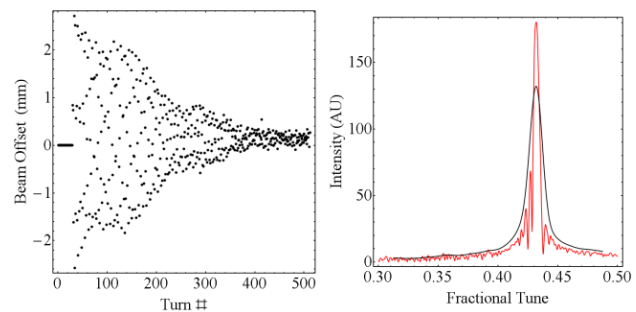

Figure 1: (Left) Example BPM data. (Right) FT (red) and smoothed FT signal (black).

*Time Domain*

A value for the tune can also be calculated in the "time domain." Here a decaying sinusoid can be fitted to the TBT BPM data:

$$Ae^{-Bx}Sin\left(\phi - \frac{Q_{x,y}x}{2\pi}\right),$$

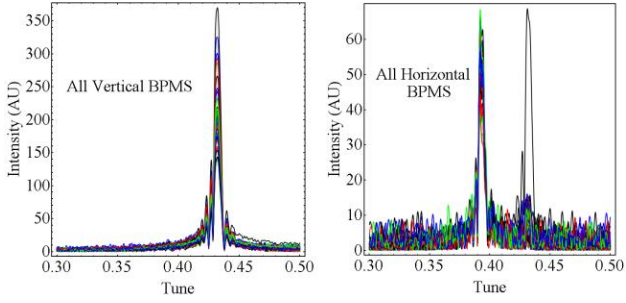

Figure 2: FT of each BPM for a particular measurement.

where x is the horizontal axis and A, B, $\phi$ and $Q_{x,y}$ are parameters to fit. Due to the coupling often a better fit could be achieved by assuming there are two frequencies present, corresponding to the horizontal and vertical tune, this can be seem in Figure 3.

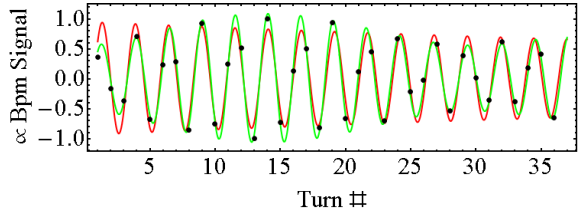

Figure 3: Single (red) and double (green) frequency decaying sinusoidal fit to BPM data (points).

## Tune Jitter

There is some jitter in the calculated tune between similar measurements. For a particular measurement each BPM gave a consistent value for the tune, as can be seen in Figure 4. A series of short experiments looking at the effect of the MI cycle of the event previous to the experiment (e.g. to see if there were hysteresis effects in the magnets) found no obvious cause for the tune jitter. Figure 5 shows the calculated tune with two different MI cycles previous to the experiment. The previous event seems to have little impact on the tune jitter. To account for this jitter many measurements were taken from which a reliable average could be taken.

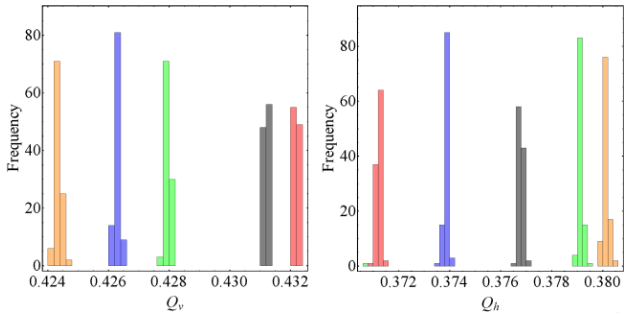

Figure 4: Tune calculated for each BPM, where different colours are different measurements with the same set-up.

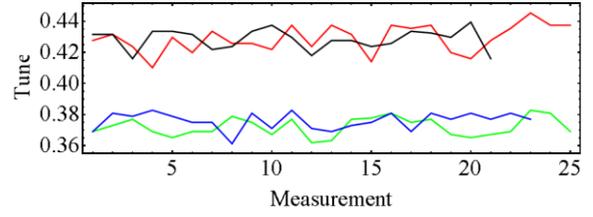

Figure 5: $Q_h$ and $Q_v$ calculated with experiment cycle placed after a *21* (red & green) or *23* (black & Blue) event.

## Difference in Time Domain and Frequency Domain Results

In general there was a slight difference between the two methods of calculating the tune, as shown in Figure 6. In this plot the tune has been calculated for multiple similar measurement and the error bars are due to the tune jitter described above. It was thought that using only a few number of turns in the TD analysis may cause the calculated difference, but after analysing all the turns using the TD method the difference was still found. For consistency and to increase the speed of calculation the rest of the analysis uses the FT method only.

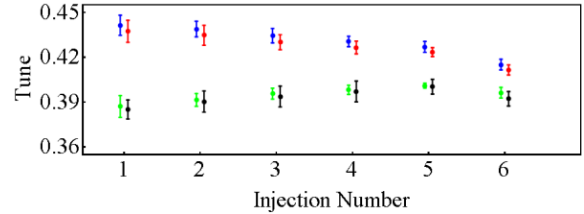

Figure 6: Mean tune of all BPM data from multiple measurements for each injection in the horizontal (green, black) and vertical (red blue) calculated using the FT (blue green) and TD (red, black) methods.

## Results

The results for each different number of booster turns are shown in Figure 7, where the errors bars represent the rms of the data.

## TUNE CHANGE WITH CHANGING RF FREQUENCY

Changing the RF frequency, $\omega_{rf}$ changes the beam momentum, resulting in a change of tune. For a fixed chromaticity, $C_\gamma$ the tune change $\Delta Q$ is:

$$\Delta Q = -\frac{C_\gamma \Delta \omega_{rf}}{\eta \omega_{rf}} \quad (1)$$

where η is the slippage factor. The tune change for the different RF frequencies is reasonably consistent for the different intensity beams, as shown in Figure 8. Averaged over all the measurements the tune change is also reasonably linear, shown In Figure 9.

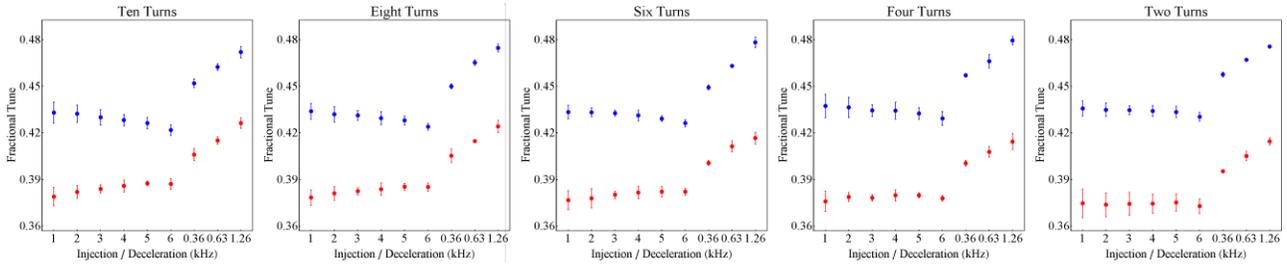

Figure 7: $Q_h$ (red) and $Q_v$ (blue). The tunes for each of the 6 injection are shown first, then, the last three points are the tunes when decreasing the RF frequency. Each plot is for a different numbers of booster turns (beam intensity).

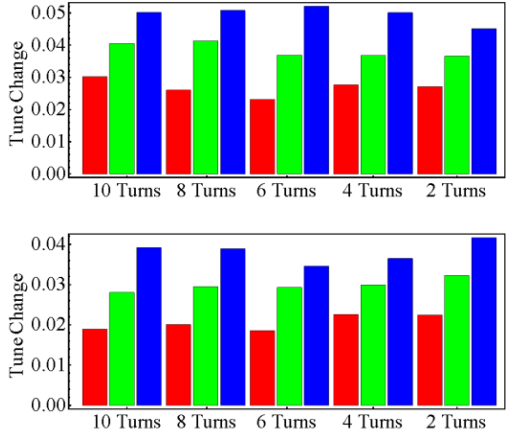

Figure 8: Vertical (top) and Horizontal (bottom) tune change for -360 (red), -630 (green) and -1260 (blue) Hz changes in RF frequency for different booster turns.

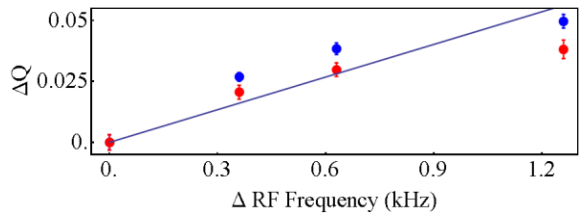

Figure 9: Vertical (red) and horizontal (blue) tune change vs RF frequency change averaged over all measurements. The line is using Equation 1.

## COHERENT TUNE SHIFT WITH INTENSITY

The tune shift due to intensity for particular bunch lengths and intensities can be described by:

$$\Delta Q_{coh} = C1 \frac{N_b}{\tau} + C2 N_b M,$$

where $N_b$ is the number of particles per bunch ($10^{10}$), $\tau$ is the full bunch length (ns) and $M$ is the number of bunches. The values for C1 and C2 for horizontal and vertical tunes in the Main injector are given in Table 1 [2]. Wall current monitor data was taken to calculate the number of particles per bunch and the bunch length [3]. From this data the measured tune shift compares well with the tune shift expected from the theory, shown in Figure 10.

Table 1: values of constants for theoretical intensity driven tune shift in the MI.

|  | Vertical | Horizontal |
| --- | --- | --- |
| **C1** | -1.390 $10^{-3}$ | -1.114 $10^{-4}$ |
| **C2** | -7.184 $10^{-6}$ | -6.915 $10^{-6}$ |

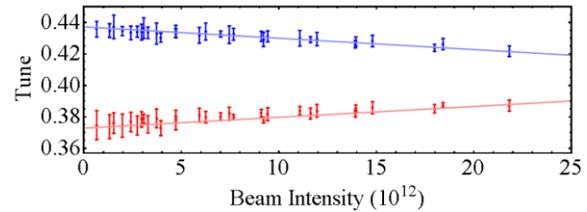

Figure 10: Expected tune vs total beam intensity (line) compared to measured tune (points) for horizontal (red) and vertical (Blue) planes.

## CONCLUSIONS

The tune has been successfully measured for six injections and then after deceleration for different intensity beams. Good agreement was found for the tune calculated from each BPM in a particular measurement. The tune changes with intensity fit well with the expected tune change from the theory. For an increase in intensity from $1\ 10^{12}$ to $25\ 10^{12}$ particles the tune change is approx. 0.015. This is as expected and gives confidence the RR can handle the expected tune changes. The tune change with RF frequency is reasonably linear and in good agreement with the expected change, although evidence for nonlinearities is apparent. The tunes jittered between similar measurements. One possible cause for this may be jitters in the quadrupole currents, as seen in the regulator values. It is not easy to check this as the regulator value needs to be converted into a current, then a magnetic field and then a tune change.

## REFERENCES

[1] Nova, Technical Design Report, 2007.
[2] Courtesy of B. Ng.
[3] D. J. Scott *et al*, "*Single Few Bunch Space Charge Effects at 8 GeV in the Fermilab Main Injector,*" these proceedings.